\documentclass[12pt]{iopart}
\usepackage{eufrak}
\usepackage{graphicx}

\begin{document}
\title{Exact equivalent-profile formulation for bent optical waveguides}
\author{D~M~Shyroki}
\address{Department of Communications, Optics and Materials,
Technical University of Denmark, Building 343v, 2800 Kgs.~Lyngby,
Denmark} \ead{ds@com.dtu.dk}

\begin{abstract}
A widespread, intuitive and computationally inexpensive method to
analyze light guidance through waveguide bends is by introducing
an equivalent straight waveguide with refractive index profile
modified to account for actual waveguide bend. Here we revise the
commonly used equivalent-index formula, ending up with its simple
extension that enables rigorous treatment of one- and
two-dimensionally confined, uniformly bent waveguides, including
tightly coiled microstructure fibers, curved ridge waveguides and
ring microresonators. We also show that such technique is
applicable only to waveguides composed of isotropic or uniaxially
anisotropic materials, with anisotropy axis directed perpendicular
to the curvature plane.
\end{abstract}

To understand and to predict light wave behavior at waveguide
bends was of major importance and interest to integrated- and
fibre-optics community from around 1970s and onwards. Today this
interest is stimulated largely by two developments: (i) increasing
the packaging density of integrated-optic circuits, backed by
minimizing integrated waveguide bend radii while keeping bend
losses at a tolerable level; and (ii) the advent of photonic
crystal fibers possessing quite complicated, high index-contrast
dielectric profiles as compared to step- or graded-index fibers
for which early theoretical methods to treat bend losses were
developed.

Numerical methods to simulate light propagation through waveguide
bends, such as beam propagation method~\cite{Rivera1995}, the
method of lines~\cite{Pregla1996, Goncharenko2005}, as well as
more general-purpose finite-element and finite-difference
techniques, can be very demanding computationally: in one recent
example~\cite{Koning2005}, a 64-processors cluster was used in
full-vector finite-element modelling. No wonder one leans to
analytic techniques to reduce computation burden whenever
possible; one such technique, perhaps the simplest and most widely
used in modelling of microstructure fibre bends
today~\cite{Baggett2003, Argyros2005, Tsuchida2005, Fini2006}, is
equivalent-profile method~\cite{Marcuse1976, Marcuse1982}. It
reduces dimensionality of the problem by replacing actual
waveguide bend by a straight piece with refractive index profile
\begin{equation}\label{StandardAlternativeEpsilon}
    n^2 = \epsilon_\mathrm{C}(1 + y/R)^2 \approx
    \epsilon_\mathrm{C}(1 + 2y/R),
\end{equation}
where $\epsilon_\mathrm{C}$ is dielectric permittivity in the bent
waveguide cross-section, as measured in Cartesian frame, $R$ is
the bend radius pointing from the $x$-directed curvature axis to a
(somewhat arbitrary)  waveguide central plane, $y$ the distance
from that plane. Designed for elegant and inexpensive treatment of
bends, though, formula~(\ref{StandardAlternativeEpsilon}) grounds
on approximations whose validity may often be questioned, as those
of weak-guidance regime and small curvature ($y/R \ll 1$), and
originally was introduced for quite a specific case of
non-magnetic, step-index, low index-contrast fibre. In this Letter
we overcome all these limitations by an alternative,
first-principle derivation of expressions for modified dielectric
permittivity $\epsilon$ and magnetic permeability $\mu$ analogous
to~(\ref{StandardAlternativeEpsilon}), but applicable to
waveguides of arbitrary cross-section under the only assumption
that bend radius is constant (while not necessarily large).

Following Schouten~\cite{SchoutenBOOK1951} and
Post~\cite{PostBOOK1962}, we start from the generally covariant
Maxwell equations
\begin{eqnarray}
    2\partial_{[\lambda} E_{\nu]} = -\dot B_{\lambda\nu}, &\quad&
    \partial_{[\kappa} B_{\lambda\nu]} = 0, \label{MaxwellInv1}\\
    \partial_{\nu} \mathfrak{H}^{\lambda\nu} =
    \dot\mathfrak{D}^\lambda + \mathfrak{j}^\lambda, &\quad&
    \partial_{\lambda}\mathfrak{D}^{\lambda} = \rho. \label{MaxwellInv2}
\end{eqnarray}
Here the square brackets denote alternation: $\cdot_{[\lambda\nu]}
= \frac{1}{2!}(\cdot_{\lambda\nu} - \cdot_{\nu\lambda})$,
$E_{\lambda}$ and $B_{\lambda\nu} = -B_{\nu\lambda}$ are covariant
electric vector and magnetic bivector fields coinciding with
electric and magnetic three-vectors $\mathbf E$ and $\mathbf B$ in
Cartesian frame, $\mathfrak{H}^{\lambda\nu} =
-\mathfrak{H}^{\nu\lambda}$ and $\mathfrak{D}^{\lambda}$ are
contravariant magnetic bivector density and electric vector
density of weight $+1$ corresponding to $\mathbf H$ and $\mathbf
D$, $\mathfrak{j}^\lambda$ and $\rho$ are electric current and
charge densities, $\lambda, \nu = 1\dots 3$. With such
transformation characteristics assigned to the electromagnetic
field quantities, Eqs.~(\ref{MaxwellInv1}), (\ref{MaxwellInv2})
are known to be form-invariant~\cite{SchoutenBOOK1951,
PostBOOK1962}, i.e., they do not change their form under arbitrary
revertible coordinate transformations. It might be practical to
convert Eqs.~(\ref{MaxwellInv1}), (\ref{MaxwellInv2}) with use of
dual equivalents $\tilde\mathfrak{B}^\kappa = \frac{1}{2} \tilde
\mathfrak{E}^{\kappa\lambda\nu} B_{\lambda\nu}$ and
$\tilde{H}_\kappa = \frac{1}{2}
\tilde\mathfrak{e}_{\kappa\lambda\nu} \mathfrak{H}^{\lambda\nu}$,
where $\tilde\mathfrak{E}^{\kappa\lambda\nu}$ and
$\tilde\mathfrak{e}_{\kappa\lambda\nu}$ are pseudo (hence tildes)
permutation fields equal to Levi-Civita symbol in any coordinate
system, to the form directly reminiscent of Maxwell equations in
Cartesian frame:
\begin{eqnarray}
    \tilde\mathfrak{E}^{\kappa\lambda\nu}\partial_\lambda E_\nu =
    -\mu^{\kappa\lambda}\dot{\tilde{H}}_\lambda, &\quad&
    \partial_{\kappa}\mu^{\kappa\lambda}\tilde{H}_\lambda = 0,
    \label{MaxwellInvMod1}\\
    \tilde\mathfrak{E}^{\kappa\lambda\nu}\partial_\lambda \tilde{H}_\nu
    = \epsilon^{\kappa\lambda} \dot{E}_\lambda + \mathfrak{j}^\kappa,
    &\quad& \partial_{\kappa}\epsilon^{\kappa\lambda} E_\lambda = \rho,
    \label{MaxwellInvMod2}
\end{eqnarray}
while the constitutive relations are implied being
\begin{eqnarray}\label{MaterialEqsInv}
    \tilde\mathfrak{B}^\lambda =
    \mu^{\lambda\nu}\tilde{H}_\nu, \quad
    \mathfrak{D}^\lambda = \epsilon^{\lambda\nu} E_\nu.
\end{eqnarray}
Here $\epsilon^{\lambda\nu}$ and $\mu^{\lambda\nu}$ are tensor
densities of weight $+1$:
\begin{equation}\label{MaterialConstantsTransform}
    \epsilon^{\lambda\nu} = |\Delta|^{-1} J_{\lambda'}^{\lambda}
    J_{\nu'}^{\nu} \epsilon^{\lambda'\nu'}, \quad
    \mu^{\lambda\nu} = |\Delta|^{-1} J_{\lambda'}^{\lambda}
    J_{\nu'}^{\nu} \mu^{\lambda'\nu'},
\end{equation}
as stipulated by transformation characteristics of the fields
$\tilde{H}_\lambda$, $\tilde\mathfrak{B}^\lambda$, $E_\lambda$ and
$\mathfrak{D}^\lambda$. We denote by $J_{\lambda'}^{\lambda}
\equiv \partial_{\lambda'}x^{\lambda}$ the Jacobian transformation
matrix, and $\Delta \equiv \det J_{\lambda'}^{\lambda}$ its
determinant. In this formulation, geometry enters Maxwell
equations (\ref{MaxwellInvMod1}), (\ref{MaxwellInvMod2}) through
material fields (\ref{MaterialConstantsTransform}) exclusively;
since the form of (\ref{MaxwellInvMod1}), (\ref{MaxwellInvMod2})
is precisely as if they were written in Cartesian components, a
multitude of analytic and numeric methods developed for
rectangular Cartesian frame apply. Furthermore, whenever
$\epsilon^{\lambda\nu}$ and $\mu^{\lambda\nu}$ happen to be
independent of one (or more) of the coordinates in a chosen
coordinate system, that coordinate (or those coordinates) can be
separated in the usual manner, like $z$ coordinate in the case of
straight homogeneous waveguide. Now we show that in cylindrical
coordinates $\{x, r, \phi\}$, such ruling out of angular
coordinate $\phi$ is the case for, in particular, isotropic guide
of arbitrary cross-section, bent along $\phi$, and specify
Eqs.~(\ref{MaterialConstantsTransform}) for that case.

\begin{figure}
  \centerline{\includegraphics[width=8.3cm]{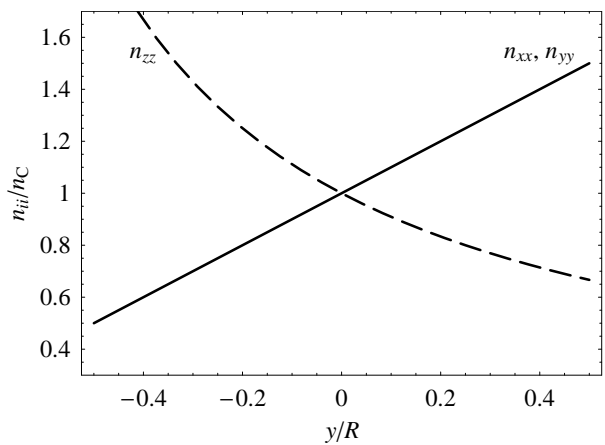}}
  \caption{Departure from refractive index $n_\mathrm{C}$ of
  waveguide with bending, for diagonal refractive tensor
  components $n_{ii} = \sqrt{\epsilon_{ii}\mu_{ii}}$ according
  to Eqs.~(\protect\ref{MyAlternativeIsotropicEps}),
  (\protect\ref{MyAlternativeIsotropicMu}).}\label{RefractiveTensorComponents}
\end{figure}

Since isotropic media are defined as those possessing scalar
(though, in general, position-dependent) permittivity
$\epsilon_\mathrm{C}$ and permeability $\mu_\mathrm{C}$, as
referred to Cartesian system, transformation rules
(\ref{MaterialConstantsTransform}) reduce to
\begin{equation}\label{isotropicMediaCartesian}
    \epsilon^{\lambda\nu} = g^{-\frac{1}{2}}g^{\lambda\nu}
    \epsilon_\mathrm{C}, \quad
    \mu^{\lambda\nu} = g^{-\frac{1}{2}}g^{\lambda\nu}
    \mu_\mathrm{C},
\end{equation}
owing to transformation behavior of the fundamental (metric)
tensor $g^{\lambda\nu} = J_{\lambda'}^{\lambda} J_{\nu'}^{\nu}
g^{\lambda'\nu'}$; its determinant $g \equiv \det g^{\lambda\nu} =
\det^2 J_{\lambda'}^{\lambda}$ insofar as $\det g^{\lambda'\nu'} =
1$ in Cartesian frame. Equations~(\ref{isotropicMediaCartesian})
resemble `effective' permittivity and permeability introduced
in~\cite{WardPendry1996}, albeit we would refrain from using the
word `effective', on the ground that
(\ref{isotropicMediaCartesian}) are nothing but transformation
rules for the permittivity and permeability of isotropic media.
With transformation from orthogonal Cartesian to cylindrical
coordinates given by
\begin{equation}\label{CartesianToCylinder}
    x = x, \quad  y = r\cos\phi, \quad  z = r\sin\phi.
\end{equation}
($x$ defines bending axis), one can find the
Cartesian-to-cylindrical transformation matrix for contravariant
components
\begin{equation}\label{Amatrix}
    J^\lambda_{\lambda'} = \left(\begin{array}{ccc}
  1 & 0 & 0 \\
  0 & \cos\phi & \sin\phi \\
  0 & -r^{-1}\sin\phi & r^{-1}\cos\phi \\
  \end{array}\right)
\end{equation}
and contravariant metric tensor
\begin{equation}\label{gTensor}
    g^{\lambda\nu} = \left(\begin{array}{ccc}
  1 & 0 & 0 \\
  0 & 1 & 0 \\
  0 & 0 & r^{-2} \\
  \end{array}\right),
\end{equation}
$g^{-\frac{1}{2}} = r$. After substituting these expressions in
Eqs.~(\ref{isotropicMediaCartesian}), separating the $\phi$
variable in Maxwell equations~(\ref{MaxwellInvMod1}),
(\ref{MaxwellInvMod2}) and differentiating the $\exp(\mathrm
i\beta\phi)$ multipliers with respect to $\phi$, normalizing the
propagation constant $\beta$ by $R$ in the usual manner and
introducing shifted coordinate $y = r - R$, one obtains
\begin{equation}\label{MyAlternativeIsotropicEps}
    \epsilon_{xx} = \epsilon_{yy} = \epsilon_\mathrm{C}(1 +
    y/R),   \quad  \epsilon_{zz} = \epsilon_\mathrm{C}(1 +
    y/R)^{-1}
\end{equation}
(we do not distinguish between co- and contravariant indices once
particular coordinate system is chosen), and similarly for
permeability components:
\begin{equation}\label{MyAlternativeIsotropicMu}
    \mu_{xx} = \mu_{yy} = \mu_\mathrm{C}(1 + y/R), \quad
    \mu_{zz} = \mu_\mathrm{C}(1 + y/R)^{-1}.
\end{equation}
Note that expressions for $\epsilon_{xx}$ and $\epsilon_{yy}$,
multiplied by their magnetic counterparts, lead to refractive
tensor components $n_{xx}^2 = \epsilon_{xx}\mu_{xx}$ and $n_{yy}^2
= \epsilon_{yy}\mu_{yy}$ precisely in line with standard
formula~(\ref{StandardAlternativeEpsilon}) in the case of
non-magnetic media. The $\epsilon_{zz}$ and $\mu_{zz}$ components
manifestly differ from the rest, however (see
figure~\ref{RefractiveTensorComponents}); this difference becomes
especially pronounced when departing from weakly guiding
approximation, as soon as non-negligible $z$-components of
electric and magnetic fields would `probe' the
$\epsilon_{zz}(x,y)$ and $\mu_{zz}(x,y)$ profiles then.

A benchmark example to illustrate the validity of
formulae~(\ref{MyAlternativeIsotropicEps}),
(\ref{MyAlternativeIsotropicMu}) is a problem of light propagation
through a homogeneous slab waveguide bend that permits precise
analytic treatment---see e.g.~\cite{Hiremath2005} for an overview.
In our approach, covariant Maxwell curl equations,
$\tilde\mathfrak{E}^{\kappa\lambda\nu}\partial_\lambda E_\nu =
-\mu^{\kappa\lambda}\dot{\tilde{H}}_\lambda$ and
$\tilde\mathfrak{E}^{\kappa\lambda\nu}\partial_\lambda
\tilde{H}_\nu = \epsilon^{\kappa\lambda} \dot{E}_\lambda$, reduce
to four independent first-order scalar equations then, grouped in
two pairs: one for the $E_x$, $\tilde{H}_\phi$ components (TE
mode):
\begin{eqnarray}
    \partial_r E_x &=& \mathrm i k\frac{\mu_\mathrm{C}}{r}
    \tilde{H}_\phi, \label{Ex}\\
    \partial_r \tilde{H}_\phi &=& \mathrm i k\left(\epsilon_\mathrm{C}r -
    \frac{n_\mathrm{eff}^2 R^2}{\mu_\mathrm{C}r}\right)E_x \label{Hf},
\end{eqnarray}
where $k$ is the free-space wavenumber, $n_\mathrm{eff} =
\beta/(kR)$ the dimensionless mode index; and another pair for
$E_\phi$, $\tilde{H}_x$ (TM mode). To derive second-order equation
for $E_x$ (which can be regarded as an eigenproblem in
$n_\mathrm{eff}$), we differentiate~(\ref{Ex}) with respect to
$r$, use~(\ref{Hf}) for $\partial_r \tilde{H}_\phi$ and $\mathrm i
k\tilde{H}_\phi = \frac{r}{\mu_\mathrm{C}}
\partial_r E_x$ to exclude $\tilde{H}_\phi$. This leads, in the regions
of constant $\mu_\mathrm{C}$, to
\begin{equation}\label{TEstandard}
    \frac{\partial^2 E_x}{\partial r^2} + \frac{1}{r}
    \frac{\partial E_x}{\partial r} +
    k^2\left(\epsilon_\mathrm{C}\mu_\mathrm{C} -
    \frac{n_\mathrm{eff}^2 R^2}{r^2}\right)E_x = 0,
\end{equation}
a Bessel equation used customarily in exact analysis of bent slab
waveguides~\cite{Hiremath2005, Jedidi2005}, while an attempt to
get similar equation with use of approximate equivalent-index
formula~(\ref{StandardAlternativeEpsilon}) fails. The results of
full-vector two-dimensional finite-difference frequency-domain
(FDFD) modelling of confined modes in a high-contrast step-index
optical fibre (figure~2) demonstrate systematic discrepancy
between approximate model~(\ref{StandardAlternativeEpsilon}) and
formulae~(\ref{MyAlternativeIsotropicEps}),
(\ref{MyAlternativeIsotropicMu}); a discrepancy which can not be
tolerated in realistic simulations of high index contrast ridge
waveguides and holey fibers, ring microresonators, or sharp fibre
bends such as those due to non-perfect alignment of fibre splices.
\begin{figure}
  \centerline{\includegraphics[width=8.3cm]{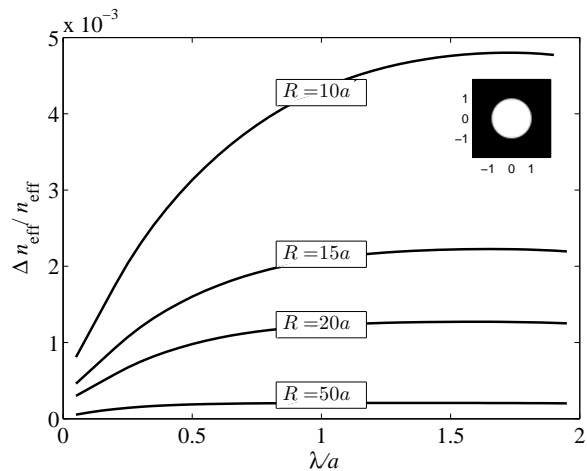}}
  \caption{The difference in fundamental mode indices
  at various bend radii $R$ of a step-index fibre of radius $a$ and
  dielectric index $n_\mathrm{core} = 1.45$ in the air background,
  calculated numerically with~(\protect\ref{StandardAlternativeEpsilon})
  and~(\protect\ref{MyAlternativeIsotropicEps}),
  (\protect\ref{MyAlternativeIsotropicMu}).}\label{SIFbending}
\end{figure}

A closer inspection of~(\ref{Amatrix}) brings to conclusion that
the only type of initial anisotropy in $\epsilon_\mathrm{C}$ and
$\mu_\mathrm{C}$ which still permits ruling out the $\phi$
dependence in $\epsilon^{\lambda\nu}$ and $\mu^{\lambda\nu}$
transformed according to~(\ref{MaterialConstantsTransform}) is
that of uniaxial crystal with anisotropy axis pointing in the $x$
direction (i.e., along the bend axis):
\begin{equation}\label{TensorialEpsilonPermitted}
    \epsilon_\mathrm{C} = \left(\begin{array}{ccc}
  \epsilon_b & 0 & 0 \\
  0 & \epsilon_a & 0 \\
  0 & 0 & \epsilon_a \\
  \end{array}\right).
\end{equation}
In that case, one gets instead of
formulae~(\ref{MyAlternativeIsotropicEps}):
\begin{eqnarray}
    \epsilon_{xx} &=& \epsilon_b(1 + y/R), \\
    \epsilon_{yy} &=& \epsilon_a(1 + y/R), \\
    \epsilon_{zz} &=& \epsilon_a(1 + y/R)^{-1},
\end{eqnarray}
and likewise for modified $\mu$. Unfortunately, however, the case
of anisotropy given by~(\ref{TensorialEpsilonPermitted}) is not
one encountered when modelling, e.g., photonic bandgap fibre
filled with liquid crystal; for that and other examples of curved
waveguides possessing anisotropy of some general kind,
\emph{rigorous} equivalent-profile formulation is not applicable
and should be substantiated by more brute-force numeric or less
stringent analytic techniques.

In summary, equivalent-profile
expressions~(\ref{MyAlternativeIsotropicEps}),
(\ref{MyAlternativeIsotropicMu}) to treat waveguide bends have
been presented. Unlike conventional
formula~(\ref{StandardAlternativeEpsilon}), they hold for
arbitrarily tight bends, arbitrary high-contrast material
profiles, and permit the inclusion of magnetic parts. This offers
three-dimensional modelling accuracy with those existing
full-vectorial two-dimensional finite-difference or finite-element
solvers which permit $3\times 3$ diagonal matrices for $\epsilon$
and $\mu$ be put in. Eliminating the restraints on bend radius and
refractive index contrast enables one to simulate also the
whispering-gallery modes in spherical, toroidal and other types of
optical microcavities. The principal limitation of the
equivalent-profile technique is that many cases of non-trivial
anisotropy can not be treated rigorously in the manner above.

\end{document}